\documentclass[11pt,oneside,letterpaper]{article}
\usepackage{amssymb}
\usepackage{amsmath}
\usepackage[dvips]{graphicx}
\usepackage{setspace}
\usepackage{fancyhdr}
\usepackage{xcolor}
\usepackage{ifpdf}
\usepackage{graphicx}
\usepackage{rotating}
\usepackage{comment}

\def\p{\partial}

\newcommand{\bi}{\begin{itemize}}
\newcommand{\ei}{\end{itemize}}
\newcommand{\bea}{\begin{eqnarray}}
\newcommand{\eea}{\end{eqnarray}}
\newcommand{\be}{\begin{equation}}
\newcommand{\ee}{\end{equation}}

\allowdisplaybreaks  

\begin{document}

\vspace*{2.5cm}
\begin{center}
{ \LARGE \textbf{A Simple Proof of the Chiral Gravity Conjecture }\\}
\vspace*{1.7cm}
 ~~~~~~~~~~~~~~~~~~~~~Andrew Strominger
\vspace*{0.6cm}
\newline
{\it Center for the Fundamental Laws of Nature\\
Jefferson Physical Laboratory, Harvard University, Cambridge, MA, USA\\}

\vspace*{0.8cm}

\end{center}
\vspace*{1.5cm}
\begin{abstract}
\noindent  Chiral gravity is three-dimensional asymptotically AdS$_3$ gravity with an Einstein, cosmological, and Chern-Simons term at 
a critical value of the coupling denoted $\mu\ell=1$.  Ordinarily, excitations of AdS$_3$ gravity are known to transform non-trivially under an asymptotic symmetry group consisting of both a left-moving and a right-moving  conformal group. However it was recently conjectured \cite{lss} that at the chiral point $\mu\ell=1$ all excitations are chiral in that they transform only under the right-moving conformal group. We show herein that at the chiral point, the group of trivial diffeomorphisms is enhanced to include the left-moving conformal transformations,  and the asymptotic symmetry group contains only one (right-moving) copy of the conformal group.  Diffeomorphism invariance then requires that all physical excitations are annihilated by left-moving conformal transformations, establishing nonperturbatively the chiral nature of chiral gravity. 
\end{abstract}

\newpage
\setcounter{page}{1}
\pagenumbering{arabic}


\onehalfspacing


Topologically massive gravity (TMG)\cite{Deser:1981wh,Deser:1982vy} with a negative cosmological constant is described by the action
\be
I_{TMG} = \frac{1}{16\pi G}\left[\int d^3x\sqrt{-g}(R+2/\ell^2)+{1 \over \mu}I_{CS}\right]
\ee
where $I_{CS}$ is the gravitational Chern-Simons action
\be
I_{CS} = \frac{1}{2\pi  }\int_{\mathcal{M}} d^3x\sqrt{-g}\varepsilon^
{\lambda\mu\nu}\Gamma^{r}_{\lambda \sigma}\left(\partial_{\mu}\Gamma^
\sigma_{{r} \nu}+\frac{2}{3}\Gamma^\sigma_{\mu\tau}\Gamma^\tau_{\nu
{r}} \right)
\ee
 and $G$ has the conventional positive sign. 
 Chiral gravity \cite{lss} is defined by taking $\mu\ell \to 1$ and imposing the standard Brown-Henneaux \cite{Brown:1986nw} asymptotically AdS$_3$ boundary conditions.  As these are local we can describe them in  Poincare coordinates for $AdS_3$
 \be\label{ads}
 ds^2=\ell^2\left({dx^+dx^-+dy^2 \over y^2}\right)
 \ee
 with the boundary at $y=0$. In these coordinates, the Brown-Henneaux boundary conditions are that fluctuations $h_{\mu\nu}$ of the metric about (\ref{ads}) fall off at the boundary according to
  \be\label{strictbc}
\left(
  \begin{array}{ccccc}
 h_{++}= \mathcal{O}({ y^0}) & h_{+-}= \mathcal{O}({y^0})  &h_{+y}= \mathcal{O}({y})  \\
 h_{-+}=h_{+-} & h_{--}= \mathcal{O}({y^0})  &h_{- y}= \mathcal{O}({y})  \\   h_{y+}=h_{+y} & h_{y-}=h_{- y} & h_{yy}= \mathcal{O}({y^0}) \\
  \end{array}
\right)
\ee
The most general diffeomorphism which preserves 
 (\ref{strictbc}) is of the form
\bea\label{bcp}
\zeta=~~&&\zeta^+\p_++\zeta^-\p_-+\zeta^y\p_y\\ =~~& [ &\epsilon^+(x^+) +{y^2 \over 2}\p_-^2\epsilon^-(x^-)+  \mathcal{O}(y^4)]\p_+ \\ ~~+ &[ &\epsilon^-(x^-) +{y^2 \over 2}\p_+^2\epsilon^+(x^+)+  \mathcal{O}(y^4)]\p_-\\~~~~~~+  & [& y\p_+\epsilon^+(x^+) +y\p_-\epsilon^-(x^-)+  \mathcal{O}(y^3)]\p_y.
\eea These are parameterized by a left and a right moving function  $ \epsilon^+(x^+)$ and $\epsilon^-(x^-)$. The subleading terms 
all correspond to trivial diffemorphisms, i.e. they have no finite surface term and hence vanish when the constraints are imposed.  The asymptotic symmetry group (ASG) is defined as the general boundary-condition-preserving   diffeomorphism (\ref{bcp})  modulo the trivial diffeomorphisms. For generic $\mu$ the ASG is generated by two copies of the Virasoro algebra.  We shall see the case $\mu\ell=1$ requires special attention.

  The consistency of these boundary conditions for generic $\mu$ was demonstrated in \cite{Hotta:2008yq}.  This demonstration involves in particular showing  that $all$ the generators (\ref{bcp}) which preserve the asymptotically AdS$_3$ boundary conditions  are well-defined expressions free of singularities as $y\to 0$ for $any$ metric of the form (\ref{strictbc}). The generator of a diffeomorphism $\zeta$ is 
\be\label{gen}
Q[\zeta] = \int_{\p\Sigma} \sqrt{\sigma}u^i T_{ij}\zeta^j\ ,
\ee
where the integral is over the $S^1$ boundary of a spatial slice $\Sigma$, $\sigma$ is the induced metric on $\p \Sigma$, $u^i$ is the timelike unit normal to $\Sigma$, and $T_{ij}$ is the boundary stress tensor.  Under Dirac brackets, the generators associated with asymptotic symmetries obey the same algebra as the symmetries themselves, up to a possible central term. The boundary stress tensor appearing in (\ref{gen}) is 
modified from that originally given for Einstein gravity in \cite{Brown:1986nw} due to the Chern-Simons term \cite{Kraus:2005vz,Kraus:2005zm,Solodukhin:2005ah,Tachikawa:2006sz}. Explicit expressions for the generators can be found in different forms in various places , the most complete discussion being  \cite{Hotta:2008yq}. We will work in an asymptotic gauge $h_{+-}=\mathcal{O}(y)$, which can always be reached with a  trivial diffeomorphism of the form $\zeta = f(x^+,x^-)y^3\p_y$.
An especially  convenient expression for the boundary stress tensor found in reference \cite{Kraus:2005zm} then reduces to 
  \bea
  T_{++}&=&(1+{1 \over \mu \ell}){1 \over 8\pi G\ell} h_{++},\\
  T_{--}&=&(1-{1 \over \mu \ell}){1 \over 8\pi G\ell}h_{--},\\
  T_{+-}&=&0.\\
  \eea
The generators  of the ASG are then 
\be
Q(\zeta)=(1+{1 \over \mu \ell}) {1 \over 8\pi G\ell}\int_{\p\Sigma} dx^+ h_{++}\epsilon^++  (1-{1 \over \mu \ell}){1 \over 8\pi G\ell} \int_{\p\Sigma} dx^- h_{--} \epsilon^-,
\ee
For chiral gravity at $\mu\ell=1$, this further reduces to the simple expression
\be\label{qz}
Q(\zeta)={1 \over 4\pi G\ell} \int_{\p\Sigma} dx^+ h_{++}\epsilon^+.
\ee
As $\epsilon^-(x^-)$ does not appear in this expression, all the diffeomorphisms parameterized by $\epsilon^-(x^-)$ are trivial, including the global $SL(2,R)_L$ subgroup.
Hence the ASG is reduced to one right-moving copy of the conformal group. Gauge invariance requires nonpertubatively that all physical excitations must be annihilated by $Q(\epsilon^-)$ and hence may carry only right moving quantum numbers. This is of course consistent with the observation  of \cite{lss}
that the left-moving energy of massless gravitons, massive gravitons\footnote{We note that the quantity denoted $h_L$ in that paper is related to the physical  left moving energy as given in \cite{Deser:2002jk,Deser:2003vh,Kraus:2005zm} by a factor involving $(1-1/\mu\ell)$.} and BTZ black holes all vanish as $\mu\ell \to 1$. 

With hindsight this result could be anticipated simply from the fact that 
$c_L=0$ for chiral gravity. If the central charge is nonzero, the associated Virasoro generators cannot be trivial, but this obstruction vanishes for the left moving sector of chiral gravity since $c_L={3\ell\over 2 G}(1-{1 \over \mu \ell})$.  It further follows from the Virasoro algebra with $c_L=0$ that the left-moving boundary gravitons, which are Virasoro descendents of the AdS$_3$ vacuum, have zero norm. Hence they should be trivially pure gauge as we have indeed seen explicitly above. 

We conclude with several comments:

\noindent (i) We wish to emphasize that while we have established nonperturbatively the chirality of chiral gravity, we have not proven that the spectrum of right-moving energies is positive or bounded from below. This remains an outstanding challenge central to the question of vacuum stability. 

\noindent (ii)It is possible that at the chiral point  (\ref{strictbc}) are not the only consistent boundary conditions. For pure Einstein gravity, (\ref{strictbc}) are known to be the only consistent possibility \cite{Brown:1986nw}\footnote{Except of course for the trivial case when the boundary conditions are so strict as to forbid all excitations \cite{Brown:1986nw}.}, and from the work of  \cite{Hotta:2008yq} this seems likely to be the only consistent possibility for generic $\mu$. However at the chiral point, since fewer components of the metric appear in the generators, it might be possible to weaken the boundary conditions and thereby get a larger ASG (e.g. an additional current algebra).\footnote{The proposal of \cite{Grumiller:2008qz} that the boundary conditions for $h_{++}$, $h_{--}$ and $h_{+-}$ be weakened from $\mathcal{O}( y^0)$ to $\mathcal{O}(\ln y)$ 
 is inconsistent in this sense because the generators (\ref{qz}) are ill-defined.  Another possiblility not so obviously ruled out is to take only $h_{--}$ to be $\mathcal{O}(\ln y)$.} Another interesting and more general possibility, not considered here, is "asymptotically non-linear" boundary conditions \cite{Barnich:2001jy,Henneaux:2004zi,Henneaux:2006hk}, in which the allowed symmetry generators are taken to depend on the asymptotic values of the  fields. 

\noindent (iii) At the chiral point, the trivial symmetry group is enlarged to include the left-moving conformal transformations. At the same time,  all left-moving degrees of freedom become trivially pure gauge.  This phenomenon of losing degrees of freedom due to an enhanced gauge symmetry appeared in a 
related AdS context in \cite{dw}. 

\noindent(iv)Following  the conjecture \cite{lss} on the chiral nature of chiral gravity, several purported counterexamples have appeared
\cite{ Carlip:2008jk,Grumiller:2008qz,Giribet:2008bw,Park:2008yy,
Grumiller:2008pr,cdww,Carlip:2008qh}. None of them take into account the enhanced symmetry at the chiral point; and 
must all fail to be true counterexamples to the chirality conjecture in one way or another. 
For example in \cite{Carlip:2008jk} linearized eigensolutions to TMG were analyzed in light cone gauge and Poincare coordinates. These eigensolutions 
are all singular on the global AdS$_3$ boundary\cite{Li:2008yz}, but it was nevertheless argued that by formation of wavepackets and a singular  gauge transformation, smooth finite energy solutions with the asymptotics (\ref{strictbc}) could be contructed. In general it is extremely difficult to analyze global properties of AdS$_3$ using both a gauge and a coordinate system which are singular on the boundary.   In \cite{Grumiller:2008qz} a non-chiral linearized solution
was explicitly found in global coordinates, but it does not obey the 
required boundary conditions (\ref{strictbc}) for chiral gravity. 
The recent paper \cite{Giribet:2008bw} claims to find a related family of nonchiral solutions which do obey (\ref{strictbc}), but as $SL(2,R)_L$  acts trivially on the physical solution space
many of these are trivial gauge transformations of one another.\footnote{This does not preclude the possibility that these modes have an interesting chiral right-moving component.}  Other constructions of purported non-chiral modes do not fully address global issues
\cite{Park:2008yy,
Grumiller:2008pr,cdww,Carlip:2008qh}.

 \section*{Acknowledgements}
This work was partially funded by DOE grant DE-FG02-91ER40654. I
wish to thank Tom Hartman, Wei Li,  Alex Maloney and Wei Song for useful conversations.


\end{document}